\documentclass{ws-procs9x6}

\newcommand{\dd}{ {\textrm d} }

\begin{document}

\title{Heavy quarks or compactified extra dimensions in the core of 
hybrid stars}

\author{G.G. Barnaf\"oldi, P. L\'evai\footnote{\uppercase{T}alk was
presented in the Workshop by \uppercase{P}. \uppercase{L}\'evai. 
\uppercase{E}-mail: \uppercase{PLEVAI@RMKI.KFKI.HU}}, \ B. Luk\'acs}

\address{KFKI Research Institute for
particle and Nuclear Physics \\
P.O.B. 49, Budapest, H-1525, Hungary }


\maketitle

\abstracts{
Neutron stars with extremely high central energy density
are natural laboratories to investigate the appearance and the
properties of compactified extra dimensions with small 
compactification radius, if they exist. Using the same formalism, these
exotic hybrid stars can be described as neutron stars with quark
core, where the high energy density allows the presence of
heavy quarks ($c$, $b$, $t$). We compare the
two scenarios for hybrid stars and display their characteristic features.}

\section{Introduction}

Neutron stars are natural laboratories to investigate the overlap of
strong, electro-weak and gravitational interaction. Many 
theoretically determined properties of these astrophysical objects
were tested by the observed properties of pulsars, and we have
detailed calculations about these stars\cite{HTWW65,Glend97,Web99,Blas01}.

However, if new perspectives appear in the description and understanding
of the gravitational interaction or in the unification of the above
interactions, then revisiting of the models becomes necessary. 
Such a reinvestigation was triggered by  the refreshed attention on
compactified extra dimensions\cite{RandSud00}. 
Extra dimensions inside neutron stars were investigated 
earlier\cite{Liddle90}, but the Kaluza-Klein (K-K) excitation modes
were not considered in the equation of state (EoS). These modes are 
important constituents of the recent gravitation theories. 
Introducing the K-K modes  into the EoS of fermion 
stars at their central core, new features  and properties 
emerged\cite{KanShir}.

Here we display a few of our ideas about these extra dimensions 
and their possible connection to particle physics. 
We summarize our numerical results
on neutron stars with different interiors (heavy quarks vs. extra dimensions)
and discuss the visibility of extra dimensions in these objects.
 
\section{The Fifth Dimension and the E\"otv\"os Experiment}

The introduction of the 5\textsuperscript{th} dimension into the real World 
has a long history. One interesting attempt is related to
the effort of Fishbach et al.~\cite{Fish86}
in finding the "Fifth Force" in the E\"otv\"os Experiment. 

Originally the E\"otv\"os Experiment\cite{Eotv22}
 has proved the Equivalence Principle
(the proportionality of inertia and gravitating mass)
with high precision. Deviation appeared in the 
9\textsuperscript{th} digit, only.
This deviation was connected to the "Fifth Force", which may be coupled
to the hypercharge $Y=B+S$ and rather weak. Considering infinite range, the
$g_W$ coupling constant of this interaction is in the order of  
	${g_W^2}/{e^2} \sim 10^{-38} - 10^{-41}$.
Such a weak force is able to explain simultaneously  
the CP violation of hyperweak interaction, which
has a terrestrial background in this interpretation~\cite{Fish86}.
This fifth force can be  a weak force disturbing the gravity measurement,
or the World is at least five-dimensional, mimicking the
existence of an extra force.

Earlier papers\cite{LukPach85,Luk02} investigated
five-dimensional geodesic motions 
assuming that metric has a Killing symmetry in the extra 
direction, $x^5$, which is space-like. 
Since  we do not observe a macroscopical 5\textsuperscript{th}
 dimension, then
$x^5$ must be compactified on a microscopic scale. 
If we cannot observe $u^5$, then we measure a false $u^i$. 
This $u^i$ satisfies an equation of motion. In lowest order the leading 
"force" term mimics Coulomb force. The specific "charge" in 
this force starts as $K_r u^r$, where $u^i$ is the true 5-velocity 
and $K^i$ is the Killing vector working in the extra direction. 
One can \mbox{obtain a constrain\cite{LukPach85}:}  
${q^2}/{Gm_0^2} < 16 \pi$, where $q$ is the charge, 
$G$ is the gravitational constant, $m_0$ is the rest mass.
Charge $q$ may have ± sign, following the sign of $u^5$, 
so 5\textsuperscript{th}
 dimension offers a possibility of geometrizing vectorial 
forces not stronger than gravity. 
There is a chance that the fifth-dimensional 
motions of particles
is connected to the quantum number hypercharge\cite{Fish86} 
or strangeness\cite{Luk02}. 

Quantization puts a serious constraint on five-dimensional motion. 
If there is an independence on $x^5$, 
then the particle is freely moving in $x^5$. 
However, being that direction compactified leads to
an uncertainty in the position with the size of $2 \pi R_c$, where $R_c$
is the compactification radius.
An angular momentum-like quantum rule appears on the "charge" connected 
to the 5\textsuperscript{th} dimensional motion, 
thus the smallest possible charge is
\begin{equation}
	{\widehat q} = n \cdot \frac{ 2 \hbar \sqrt{G}}{cR_c} \ \ .
\end{equation}	
Because of the extra motion into the fifth
dimension, an extra mass appears in 4D descriptions. Considering
$R_c \sim 10^{-12}- 10^{-13}$ cm, together with the extra 
interaction in the range of ${\widehat q}$, this extra 
"mass" is ${\widehat m} \sim 100$ MeV. 

\section{Strange Compact Stars and the Cyg X-3} 

We do know that pulsars are very compact objects with mass 
$M \sim 1 \  M_{\odot}$ in solar mass unit and  radius $R \sim 10$ km.
Neutron matter in its own gravity can produce such configurations, 
thus the simplest explanation of the pulsars is connected to
neutron stars produced  in gravitational collapse\cite{HTWW65,Glend97}. 

The observed maximal mass for pulsars is $\sim 1.5$ $M_{\odot}$.
Early theoretical calculations\cite{HTWW65} 
with non-interacting one component
neutron matter have yielded to maximum mass  $\approx 0.67 \  M_{\odot}$.
The introduction of nuclear interaction among neutrons\cite{Glend97}
can increase this mass by a factor of 2, however details may be still 
crucial, because our knowledge is very much limited
about the properties of interactions at 10 times of normal 
nuclear densities. This leads to the conclusion, that
some heavy pulsar may not be a simple neutron star, but a more complicated
object with an exotic core\cite{GlendFaro}. 
Further astrophysical observations strengthen this expectation.

In 1987 two disjoint neutrino bursts were measured from the supernova (SN)
1987A, separated by several hours.
Various SN models predict one neutrino burst when/if the neutron star 
is formed, but never two. One could understand the double burst 
with the formation of two possible compact star configurations:
hyperon star containing strange hadrons heavier than neutron or
quark star containing a deconfined quark matter  core.

Between 1981 and 1991 strong muon showers were detected with a 4.8 hours
periodicity\cite{Thomas91}. 
The direction of these showers is that of the Cyg X-3 object, 
which is a close binary object:
one component is a normal massive star with $4 \  M_{\odot}$, the other 
is a compact star with orbital period of 4.8 hours\cite{Klis81}
and the muons have shown the same periodicity\cite{Mars85}. Their 
distance from Earth is $\approx 40,000$ ly. Cyg X-3 is a very intensive 
source in a wide spectral range and surely one important  phenomena
is the impact of stellar wind on the surface of the compact component
generating particle packages, which will hit the Earth later.
These packages must consist of neutral particles traversing the 
40,000 light years, otherwise galactic magnetic fields would have 
smeared away the 4.8 hour period. However, photons or neutrinos 
would generate far too few muons in the terrestrial atmosphere\cite{Mars85}. 
So dilambdas\cite{Baym85} or (so far undiscovered) strange nuggets
were suggested to be the messengers, which can be stable\cite{Baym85,Fleck89}
 and Cyg X-3 B was identified
to be a hyperon (or a strange quark) star generating these strange
particle packages in its surface. 
Theoretical neutron star calculations have found such a strange quark
star configuration\cite{Kett95}, which  was stable. Such an object could
be the source of these strange messengers.

\section{ The Tolman-Oppenheimer-Volkov Equation} 

The formation of neutron stars is based on the existence of a hydrostatic 
equilibrium between strong and gravitational interaction.  
In 4D a static spherically symmetric fluid configuration 
has to satisfy 3 nontrivial components of the Einstein 
equation. 
Two of them give pure quadratures, the 
third constrain leads to the Tolman-Oppenheimer-Volkov 
equation\cite{HTWW65,Glend97,Opp39}:
\begin{equation}
	\frac{\dd p}{\dd r} =  
- \frac{[p(r)+\varepsilon(r)][G  M(r)+
4 \pi \ G \ r^3 \ p(r)]}{r[r-2 \ G  M(r)]} \ \ ,
\label{TOV}
\end{equation}
where $p(r), \ \varepsilon(r)$ are the radial distribution of the 
pressure and the energy density, and
$M(r)$ is the mass of the neutron star within radius $r$.

Before solving eq.(\ref{TOV}) we have to specify the EoS of the interior
matter in the form: $p = p(\varepsilon)$.  Since the configuration is 
static, we can take cold matter 
in equilibrium and considering the $n$ density of some conserved particle
number (e.g. baryon number for neutron): $\varepsilon = \varepsilon (n)$.
Applying thermodynamic identities the pressure can be extracted:
$p = n \ \dd \varepsilon / \dd n - \varepsilon$. 

In eq.(\ref{TOV}) the central energy density, $\varepsilon_{cent}$, 
can become the initial condition.  Starting from a homogen central core with
radius $R_0 \sim 1$ cm, one can integrate the TOV equation 
until the surface. 
Because the static interior solution needs to match the exterior
vacuum condition on this surface, the $p=0$ condition will fix the
radius at $R\sim 10$ km.
Finally, the equilibrium configurations depend on a single parameter,
$\varepsilon_{cent}$.

For stability criteria see Ref.\cite{HTWW65}, where this question 
is investigated in details. 
Practically: stability can change only at extremum points of the 
curve $M(\varepsilon_{cent})$, where either the smallest real 
eigenfrequency of oscillation becomes imaginary through 0, 
or vice versa. The actual change 
depends on the sign of $\dd R/\dd \varepsilon_{cent}$ at that point 
(see Chap. 6 and 7 in Ref.\cite{HTWW65} ).

In case of quark stars a first order phase transition can appear inside 
the star. We obtain a critical surface at the critical pressure, $p_{cr}$,
where two different $\varepsilon$ values can be found, depending on
the EoSs.  Since $p$ is continuous, then the sectionwise solutions of the 
TOV equation can be determined. However, 
the solution of the TOV equation jumps from $\varepsilon_1$ to 
$\varepsilon_2$.
This jump is absent in the case of second order phase transition.

Introducing extra dimensions, in the general case 
the 5-dimensional Einstein equations 
cannot be reduced into a single, TOV-type first order 
differential equation, but a coupled differential equation 
system is  generalized\cite{Liddle90,KanShir}.
Assuming a radially independent compactification circumference
the TOV equation reappears with a 5-dimensional interpretation. 
The details of this correspondence is discussed in the Appendix.

\section{Neutron Star in 4 Dimensions} 

Our reference object is an "ideal" neutron star, 
which consists of pure neutron matter.
We use free fermion gas EoS with multiplicity
$d_N=2$. We neglect the "normal matter" constituent,
which is a very thin and negligible layer at the surface of neutron stars.
The radius is obtained in the range of $R\sim 5-15$ km for
$\mu_N > 1$ GeV, reproducing earlier results from Ref.\cite{HTWW65}.
Real neutron star calculations are more complicated
(see Refs.\cite{Glend97,Kett95}), however we want to demonstrate some features 
and the above simple model is just appropriate for this task.

On Figure 1  the mass and the radius of the 
neutron star are displayed as the function of central energy density.
The first upgoing part of $M(\varepsilon_{cent})$ is stable until
the star ($\star$), the next  
is unstable, and all downslope parts of $M(\varepsilon_{cent})$ are unstable. 
In parallel, we constructed the $M(R)$ curve (see right hand side),
where the spiraling behaviour can be clearly identified\cite{HTWW65}.

\vspace*{-8mm}
\begin{figure}[htb]
\begin{minipage}[t]{6.5cm} {\epsfxsize=6cm \epsfbox{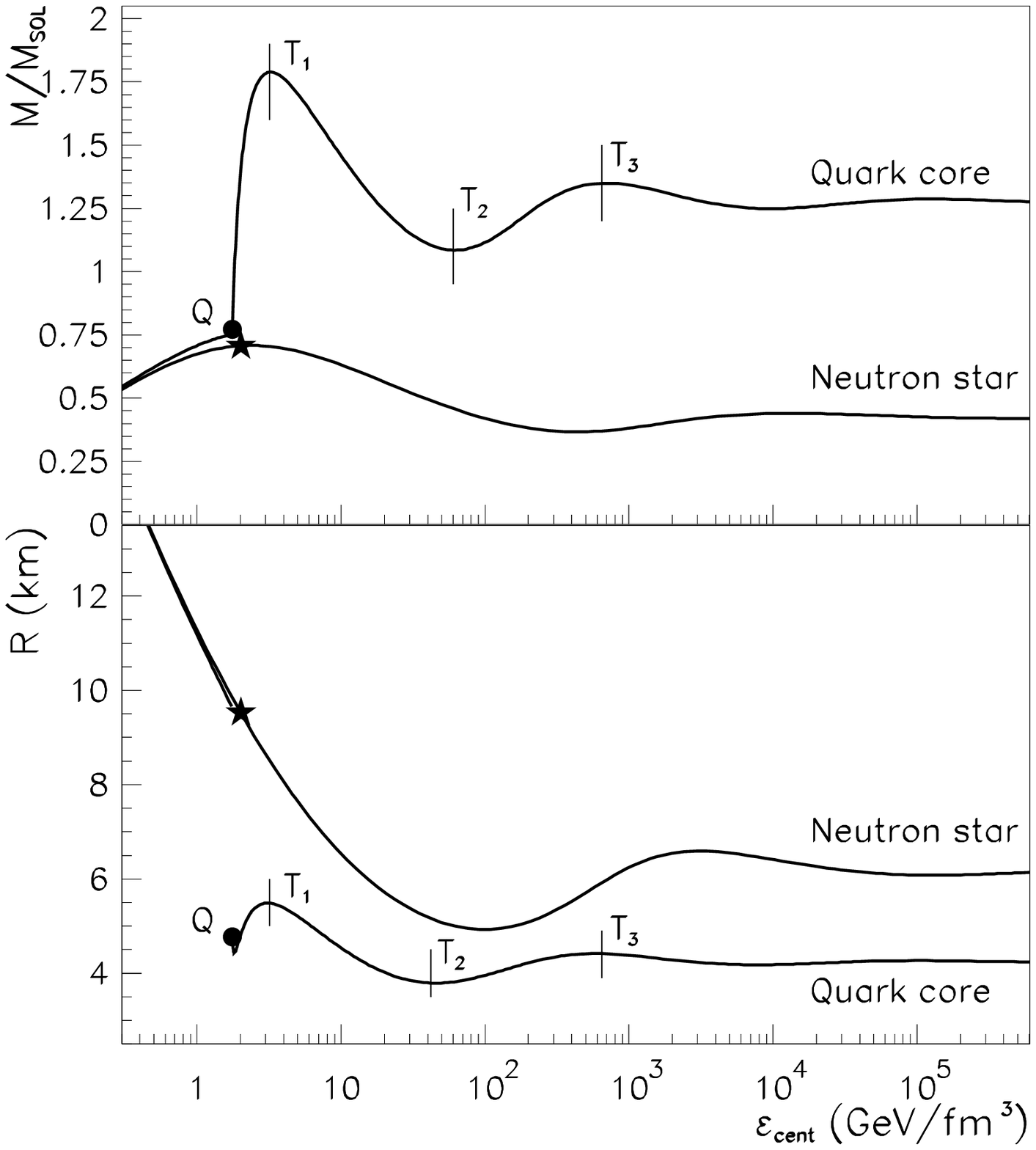}}
\end{minipage} \hfill
\vspace*{- 67mm} \hspace*{6.0cm} \hfill
\begin{minipage}[t]{6.5cm}
{\epsfxsize=6.0cm \epsfbox{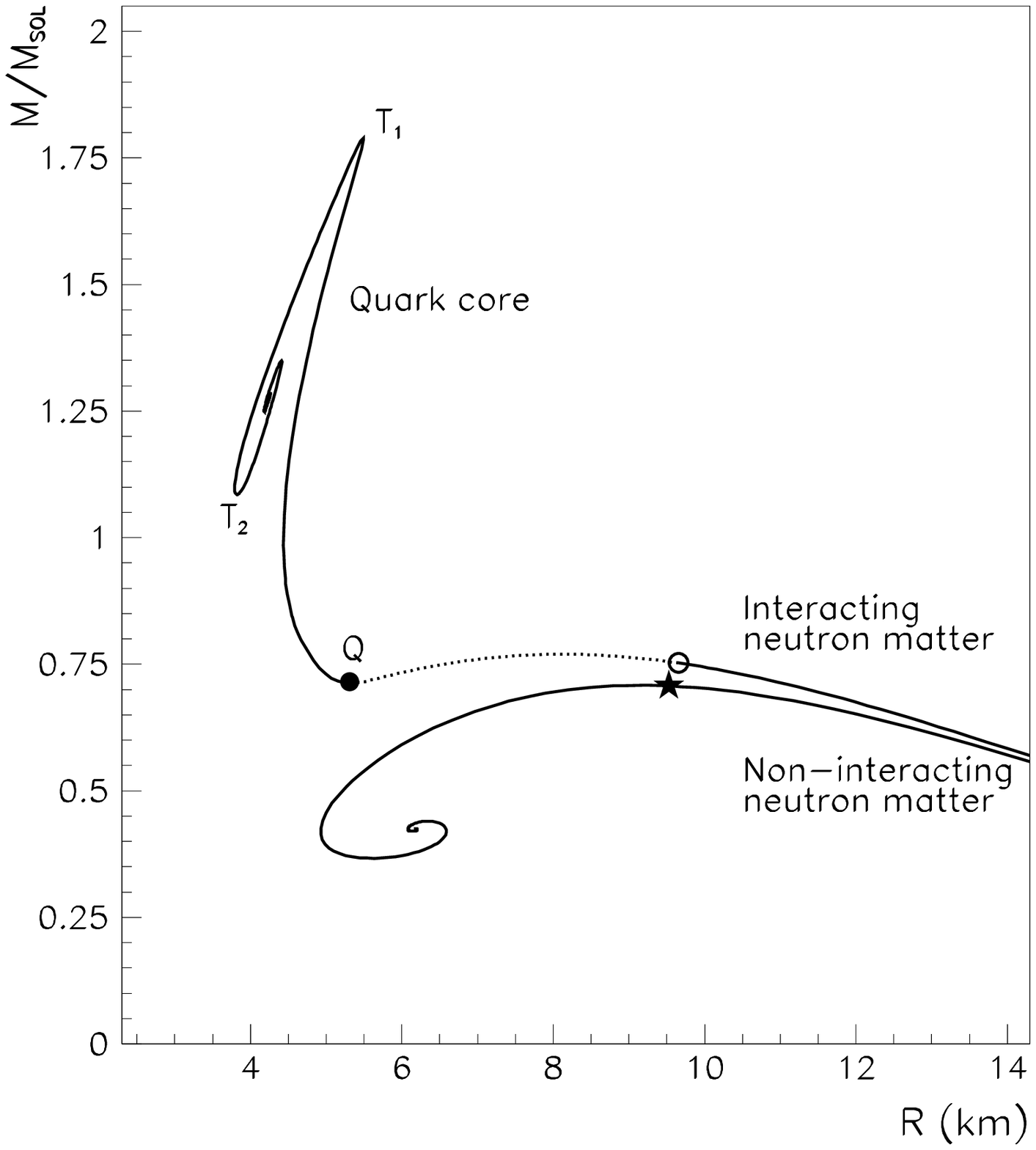}}
\end{minipage}
\vspace*{-8mm}
\caption{The mass and the radius of neutron stars containing 
pure neutron matter  or light quark core.
The $M(R)$ function is displayed on the right hand side for
both cases.  
Dotted line between full dot and open dot indicates
a discontinuity before the appearance of  quark core
($Q$).
The $T_1$, $T_2$, $T_3$ display turning points for quark star.
\label{fig1}}
\end{figure}

\vspace*{-4mm}
Introducing a quark core with light ($u$ and $d$) quarks,
a first order phase transition occurs inside the neutron star.
Considering an interacting EoS for the neutron matter\cite{BLL03}  and 
a free fermion gas EoS for the quark matter with a bag constant 
$B=0.25$  GeV/fm$^3$, the quark core appears  ($Q$) at
$\varepsilon_{cent}=1.7$ GeV/fm$^3$. No stable configuration was found between
turning points $T_1$ and $T_3$, e.g. light quark star is unstable around $T_2$.
The region between $Q$ and $T_1$ is promising, but 
a more sophisticated model is needed.

\newpage

\section{Heavy Flavours in Hybrid Stars} 

One can observe particles heavier than the neutrons, 
carrying conserved quantum numbers, namely the
$S$ strangeness, $C$ charmness, $B$ bottomness (and maybe $T$ topness). 
Weak interaction is allowed to create these flavours inside 
the core of a neutron star at extremely high energy density.

Let us construct an {\it "ideal heavy hadron star"} consisting of neutrons and
other heavier hadrons with these heavy flavours. For simplicity 
we consider the neutral
$\Lambda_S(1115)$, $\Lambda_C(2452)$ and $\Lambda_B(5624)$ 
(we stop at central chemical potential $\mu_{cent}=25$ GeV, 
thus we miss the $\Lambda_T$). The
multiplicities are $d_i=2$ and free fermion EoS is used in the
TOV equation as previously.

Figure 2 displays our results for heavy hadron star,
which are very close to the pure neutron star case,
e.g. the stability and the spiraling feature remain the same.
Just before the first peak stable configuration  
appears with $\Lambda_S$ core and $N$ mantles 
({\it "hyperon star"}~\cite{Glend97}). 
In the first maximum ($\star$),
far before the appearance of $\Lambda_C$,
the stability ceases and never returns.

\vspace*{-8mm}
\begin{figure}[htb]
\begin{minipage}[t]{6.5cm} {\epsfxsize=6cm \epsfbox{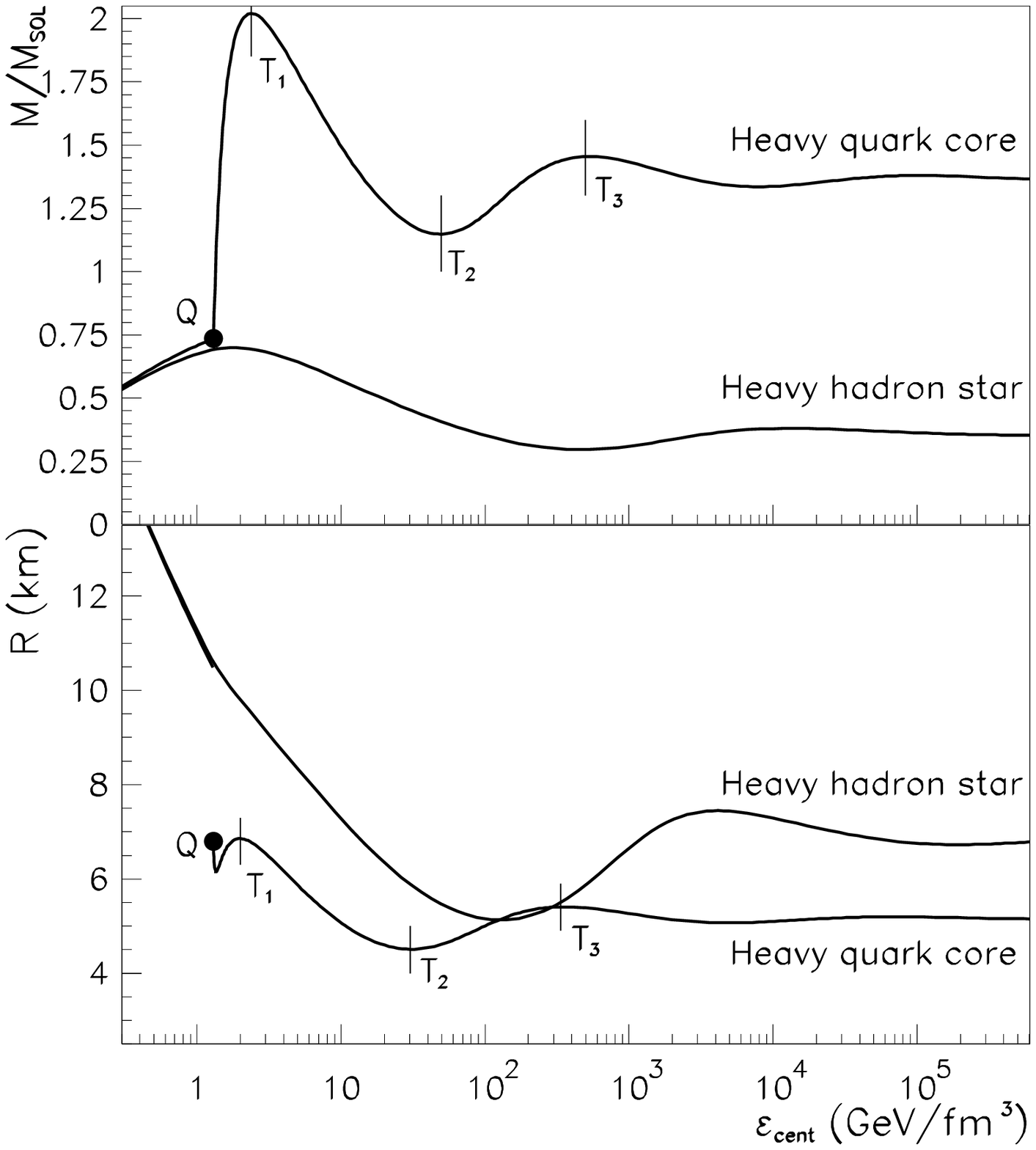}}
\end{minipage} \hfill
\vspace*{- 67mm} \hspace*{6.0cm} \hfill
\begin{minipage}[t]{6.5cm}
{\epsfxsize=6.0cm \epsfbox{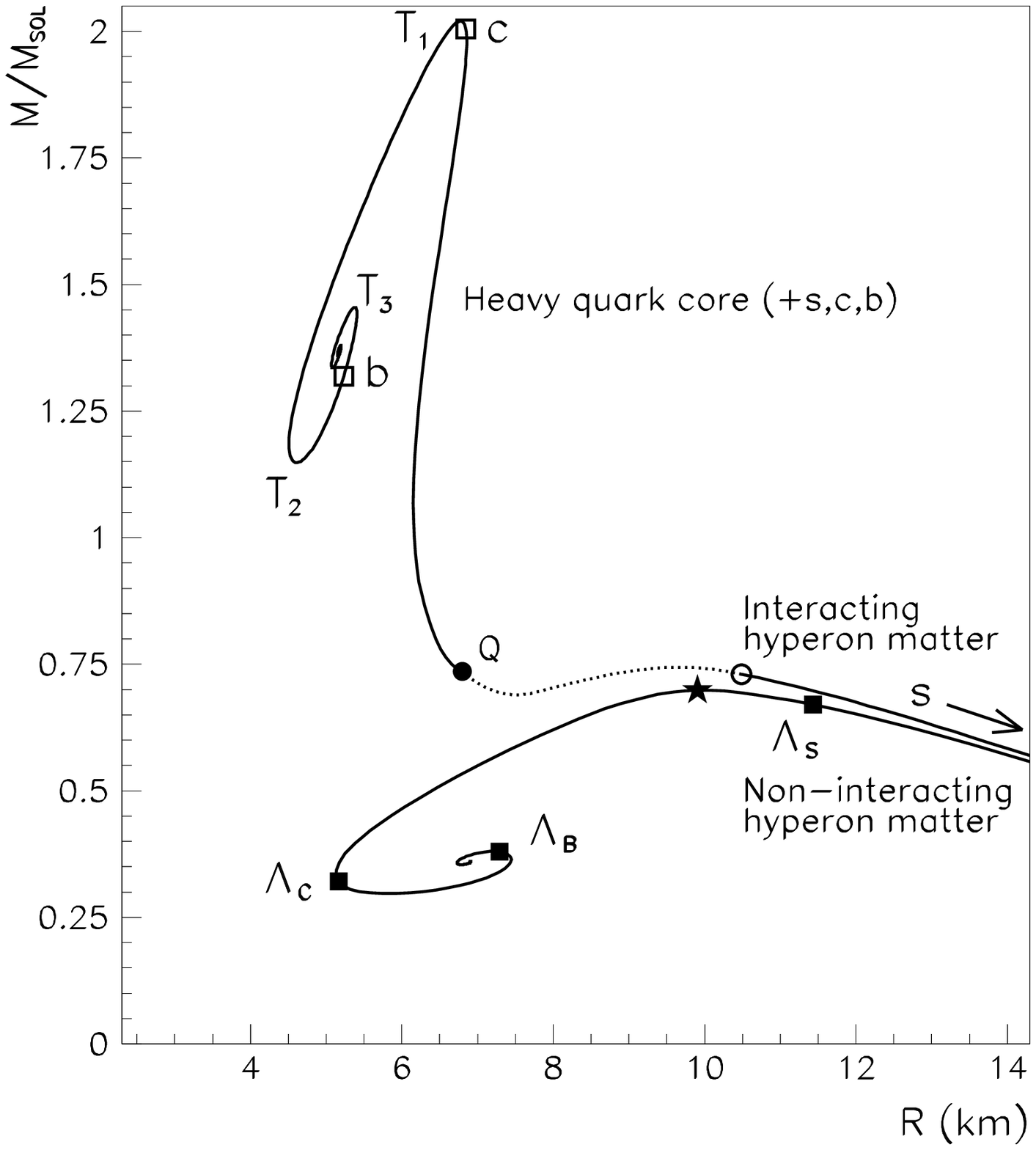}}
\end{minipage}
\vspace*{-8mm}
\caption{The mass and the radius of heavy hadron stars 
with and without heavy quark core.
The $M(R)$ function is displayed in the right hand side for
both cases.
\label{fig2}}
\end{figure}

\vspace*{-4mm}
Introducing a quark core with light ($u,d$) and heavy ($s,c,b$) quarks,
a first order phase transition have to be considered, again.
Since the $s$ quark is relatively light, then some modifications appear, 
but the investigated features remain similar.
The quark core appears ($Q$) at $\varepsilon_{cent}=1.3$  GeV/fm$^3$,
when $u,d,s$ quarks are already present. 
No stable configuration was found between
turning points $T_1$ and $T_3$, but close to $Q$  stable states
({\it "strange quark stars"} \cite{Kett95}) may appear.
However, no charm, bottom or top quark stars are expected
to be stable at high central energy densities\cite{Prisz94}.

\section{Hybrid Stars in 5 Dimensions} 

Now let us consider a neutron star with excitations in the
5\textsuperscript{th} dimension, which are compactified. In this case neutrons
are moving into the direction of $x^5$. 
In the special case of $x^5$-independence this motion appears
as an extra degrees of freedom with larger masses:
\begin{equation}
(m_N^{(i)})^2 = \left( {i}/{R_c} \right)^2 + m_N^2 \ . 
\label{m5}
\end{equation}
Here the integer $i$  means the excitation level,  $R_c$ is
the compactification radius of the 5\textsuperscript{th} dimension
and $m_N^{(0)} \equiv m_N(940)$.
The EoS is generated as the superposition of the different 
K-K modes\cite{KanShir}, summing up the free fermion EoS 
at masses from eq.(\ref{m5}). If we choose $R_c=0.33$ fm, 
then the first excitation has the mass identical with that of the 
$\Lambda$ particle, $m_N^{(1)} = \Lambda(1115)$.
The TOV equation remains valid to find the appropriate
equilibrium states.

Figure 3 displays the (non-interacting) heavy hadron matter case (full line)
and the 5-dimensional cases with two different compactification radii,
$R_c=0.33$ fm (dashed line) and $R_c=0.66$ fm (doted line).  
The filled and open triangles indicate the K-K modes.

\vspace*{-8mm}
\begin{figure}[htb]
\begin{minipage}[t]{6.5cm} {\epsfxsize=6cm \epsfbox{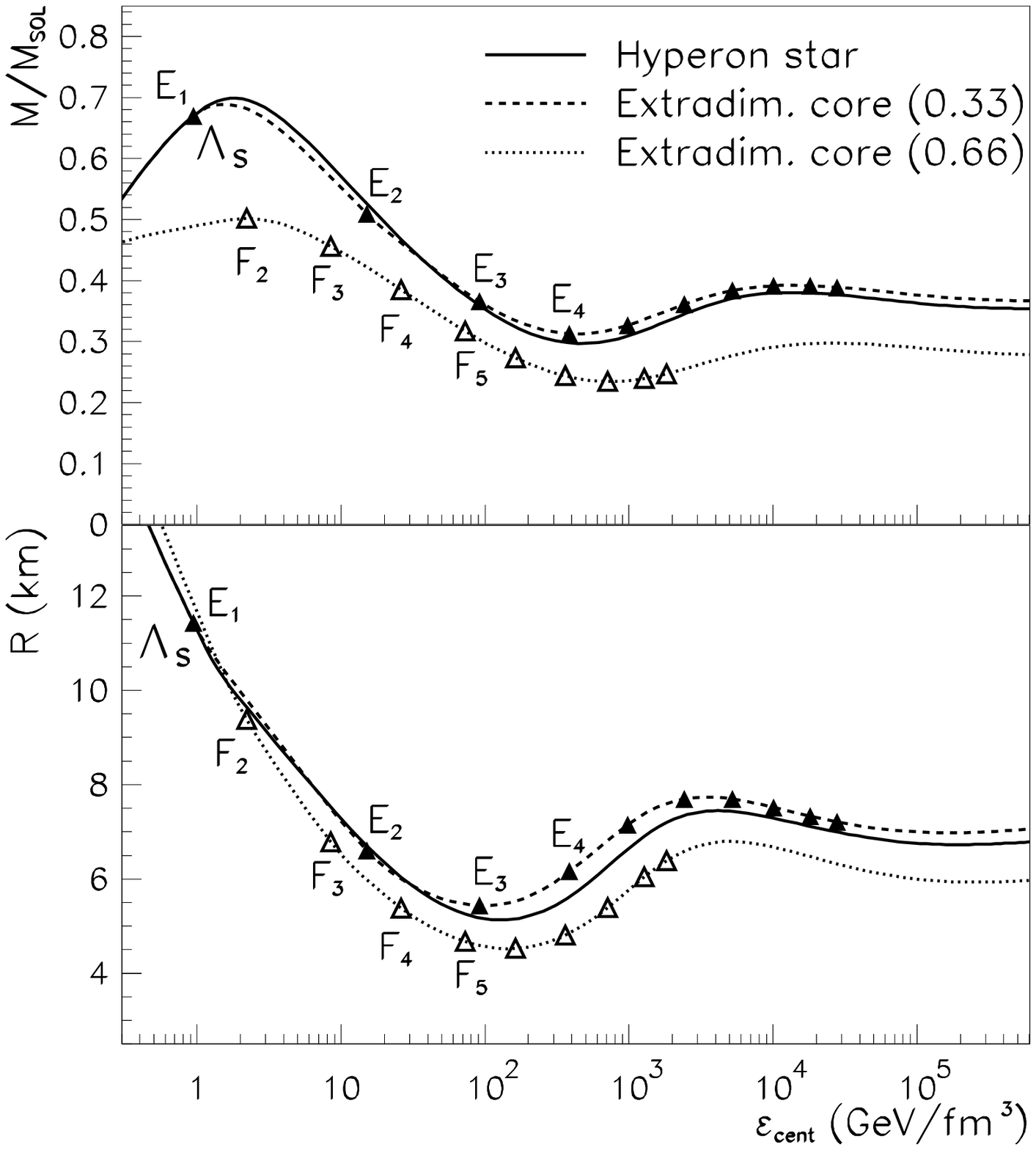}}
\end{minipage} \hfill
\vspace*{- 67mm} \hspace*{6.0cm} \hfill
\begin{minipage}[t]{6.5cm}
{\epsfxsize=6.0cm \epsfbox{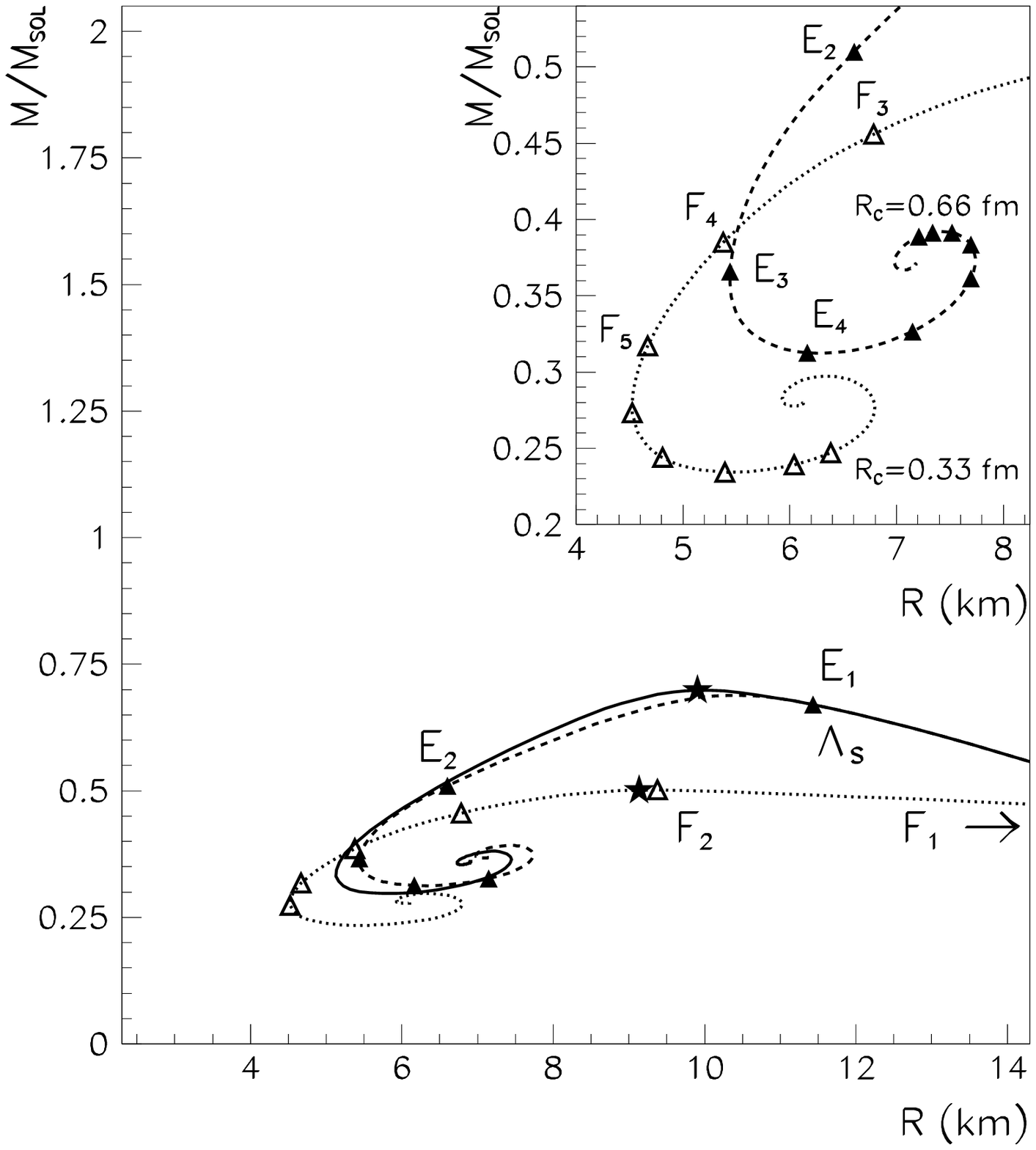}}
\end{minipage}
\vspace*{-8mm}
\caption{The mass and the radius of heavy hadron stars in 4 dimensions
and neutron stars including extradimensional K-K modes into the
core.
\label{fig3}}
\end{figure}

\vspace*{-4mm}
In the case of  $R_c = 0.33$ fm the first K-K mode ($E_1$)
appears in a stable configuration indicated until the star ($\star$). 
Choosing $R_c=0.66$ fm, we have $F_1$ and $F_2$ modes before the end of 
stability (the star ($\star$))  on the dotted line.
In the case of $R_c > 0.25$ fm,
one or more extradimensional modes can fit into the core before
loosing the stability of the hybrid star.

\newpage

\section{More Extra Dimensions} 

The introduction of the 5\textsuperscript{th} dimension is
a minimal extra-dimensional model. One may assume an
$x^6$, or more extra dimensions,  roughly on the scale of $x^5$. 
(The idea of six-dimensional microphysics  goes back to 
35 years\cite{AhnAnd70}.) 
The opening of the 6\textsuperscript{th} dimension leads to similar
results as displayed in Figure 3, however now more combinations of
extra-dimensional excitations may appear in the stable region.
In Ref.\cite{KanShir} even two stable regions and two maximums on
$M(R)$ appeared, indicating the complexity of the higher dimensional
results. It is interesting to mention, that the existence of 
such a second stable peak can explain the double neutrino burst 
of SN 1987A.  However, detailed calculations are needed
to verify this explanation.

\section{Conclusions}

We have demonstrated in a simple model for compact star that
neutron stars with hyperon or extradimensional core
are very similar objects. The TOV equation leads to similar structure
with well-defined stability region, where the lowest K-K modes may appear
in the extradimensional case. The main reason is connected 
to the size of the compactification dimension: 
assuming $R_c=0.33$ fm one obtains
the mass of $\Lambda^0$ for the first K-K mode, which dominates
the inner structure of the hybrid star in the stable region. 
We saw in Section 2 that motion in the fifth dimension results in an apparent 
new quantum number, similar in structure to strangeness,
so a neutron moving  into the 5\textsuperscript{th} dimension 
may be seen as a $\Lambda_S$ particle.
(The resulting apparent "violation of equivalence" will appear
roughly in the order of Ref.\cite{Fish86} as shown in Ref.\cite{LukLad93}.)
Now the dilambda explanation of  muon bursts
triggered by Cyg X-3 for short 
periods  is not utterly hopeless,
although rather cataclysmic events  are needed
to get strange matter to the surface.

The introduction of more extra dimensions leads to the 
appearance of further stability regions and the existence of more than one
stable hybrid star configuration.  Such a result is supported by the
double neutrino shower arrived from SN 1987A.

The investigation of neutron stars with light and heavy quark core
leads to different characteristics and to higher star masses. It remained
open the question of stability of these objects, 
more sophisticated models\cite{Blas01a,Blas01b,Blas03}
are needed for interacting neutrons and quarks. 
Interesting question is
if any extra-dimensional setup is able to mimic the features of the
heavy quark stars.
The connection between
strange quark and extra-dimensional propagation of light quark
deserves further studies.

\newpage
\section*{Appendix A: 5D Static Equilibrium and the TOV Equation} 

Let us consider a spherical equilibrium situation in 5 dimensions. 
The solution of the Einstein equation must be stationary and 
spherically symmetric. Since we do not have information about the 
dependence on the microscopic, compact dimension, and cannot observe either, 
let us assume $x^5$ independence.
Thus we have an $U(1)\otimes SO(3) \otimes U(1)$ Killing symmetry, 
5 Killing vectors altogether, with 4-dimensional transitivity; 
SO(3) is transitive in 2 dimensions. Using permitted coordinate transformations,
on the analogy of the derivation of Ref.~\cite{HawEll73}
we arrive at the general form of the line element:
\begin{equation}
	\dd s^2 = g_{00} \dd t^2 + g_{11} \dd r^2 - r^2\dd \Omega^2 + 
g_{55} \dd x^5 \dd x^5 + 2g_{05}\dd t \dd x^5 \ \ .
\end{equation}
Here $\dd \Omega^2$ is the usual spherical elementary surface.
The still unknown components $g_{11}$, $g_{55}$ and $g_{05}$ 
are independent on $t$, 
$x^5$ and the angles. Furthermore $g_{11}, \ g_{55} <0$. 
However $g_{05}$ cannot be transformed away generally.

If we require static temporal symmetry instead of mere stationary, 
then the timelike Killing vector will be hypersurface-orthogonal and 
even $g_{05}$ can be transformed away. Here we are content with this subclass, 
since in 4 dimensions all stationary spherical 
fluid solutions were static too \cite{HawEll73,Perj76}. 
The more general stationary class, with some interplay or drag between $\dd t$ 
and $\dd x^5$ would deserve further attention in a subsequent paper.

Now we are confronted with a further choice. 
We are restricted to static fluids, and by definition, the 3-dimensional 
stress tensor of a static fluid is isotropic (Pascal law). 
However, we are in 5 dimensions, so the space-like section is of 
4 dimensions. What is the proper generalisation of a fluid now?

The problem is discussed but by no means obligatorily solved in 
Ref.~\cite{Shah}. 
For a macroscopical 5\textsuperscript{th} dimension full equipartition, 
(thus 4 dimensional isotropy) would have good arguments. 
However, in our case the 5\textsuperscript{th} dimension 
is microscopic, when usual interactions in fluids (e.g. van der Waals forces) 
could not establish full spatial isotropy. Until a convincing ansatz 
is found, we decouple the 55 component of the Einstein equation. 
The remaining equations have always two sets of solution:
either $\dd g_{55}/ \dd r = 0$ or not. 
In the first case the equations can be converted into the TOV equation,
and $T_{55}$ can be calculated afterwards.
In the second case a system of coupled differential equations appears
and the compactification radius has 
radial dependence\cite{KanShir}.

\newpage

\section*{Acknowledgments}
P.L. thanks for the warm hospitality of A. Krasnitz and the 
Workshop organizers,
they have created a wonderful meeting at Faro.
This work was supported by the OTKA T032796, T034269 and T043455.

\end{document}